\documentclass{ringb99}
\usepackage{graphics}
\newcommand{\HI}{{\rm H{\sc I\,}}}
\newcommand{\HII}{{\rm H{\sc II\,}}}
\newcommand{\HeI}{{\rm He{\sc I\,}}}
\newcommand{\HeII}{{\rm He{\sc II\,}}}
\newcommand{\HeIII}{{\rm He{\sc III\,}}}
\newcommand{\unit}[1]{\ifmmode {\rm\ #1\,} \else {$\rm #1$} \fi}
\newcommand{\angstrom}{\unit{\AA}}

\newcommand{\etal}{et al.\,}
\newcommand{\Halpha}{\unit{H\alpha\,}}
\begin{document}
\title{A reanalysis of EUV emission in clusters of galaxies}
\author{Stuart Bowyer\inst{1} \and Thomas W. Bergh\"ofer\inst{2} \and Eric 
Korpela\inst{1}}
\institute{Space Sciences Laboratory, University of California, Berkeley, CA
  94720-7450, USA
\and
Max--Planck--Institut f\"ur extraterrestrische Physik,
 Giessenbachstra{\ss}e, 85740 Garching, Germany
}
\maketitle

\begin{abstract}

We report a new analysis of diffuse EUV emission in clusters of galaxies.
We find the cluster emission is strongly influenced by the variation of the 
telescope sensitivity over the field of view and upon the details of the 
subtraction of the EUV emission from the X-ray plasma. We investigate these 
effects on Abell 1795, Abell 2199, and the Coma cluster. When we use the
appropriate correction factors, we find there is no evidence for any excess 
EUV emission in Abell 1795 or Abell 2199. However, we do find extended EUV 
emission in the Coma cluster and in the Virgo cluster using our new analysis 
procedures, confirming that in at least these clusters some as yet 
unidentified process is operative.

\end{abstract}

\section{Introduction}
Extreme ultraviolet (EUV) emission in excess of that produced by the 
well-studied X-ray emitting gas in clusters of galaxies has been reported in 
five clusters of galaxies. The effective bandpass of the EUVE telescope 
employed in these observations is defined by the intrinsic response of the
telescope combined with the absorption of the intervening galactic interstellar
medium (ISM). This bandpass has a peak at 80\AA with 10\% transmission at 66 
and 100\AA. A variety of instrumental effects that might have explained these 
results have been advanced but a detailed analysis has shown these factors 
cannot explain the data (Bowyer, Lieu \& Mittaz, l998). 

It is interesting to note that the EUV excess is detected in some ROSAT
images.  However, the effect is sufficiently marginal that the ROSAT results
can almost be explained away through the use of particular combinations of
intervening Galactic ISM and its ionization state, and different cross
sections for absorption by hydrogen and helium (Arabadjis \& Bregman, 1999).
The EUVE results, however, cannot be explained in this manner.  It is also
interesting to note that the EUV excess has been reported in every cluster
investigated with EUVE.

A number of suggestions have been made as to the source of this EUV emission.
Initial work focused on additional components of ``warm gas'' ($\sim 10^6$\,K).
The problem with this suggestion is that gas at this temperature is near the
peak of the cooling curve and substantial energy is needed to supply the
energy radiated away.  One mechanism that can provide this energy is
gravitational condensation.  Cen \& Ostriker (1999) have suggested that a
pervasive warm intergalactic gas constitutes the majority of matter in the
Universe; as this gas coalesces onto clusters of galaxies, it could produce
the energy needed to sustain the EUV emitting gas.

Several authors (Hwang, 1997; En{\ss}lin \& Biermann, 1998) have suggested the
EUV flux in the Coma Cluster is inverse Compton (hereafter: IC) emission
produced by the population of electrons producing the radio emission
scattering against the 3$^{\circ}$\,K Black Body cosmic background.  However,
Bowyer \& Bergh\"ofer (1998) have shown that the existing population of radio
emitting cosmic ray electrons cannot be responsible for the EUV emission in
the Coma cluster, and some other population of cosmic rays will be required if
this mechanism is the source of the EUV emission in this cluster.  Lieu et al.
(l999a) have suggested that the Coma cluster contains a large population of
cosmic rays which are producing the 25 to 80 keV emission seen by BeppoSAX
(Fusco-Femiano et al., 1999) and RXTE (Rephaeli, Gruber \& Blanco 1999) via 
IC emission. They propose this population of
cosmic rays extrapolated to lower energies will produce the observed EUV flux
by IC emission. However, these authors have not addressed the fact that this
population of electrons will produce a spatial distribution of the EUV flux
which is inconsistent with the observational results of Bowyer \& Bergh\"ofer
(1998).

En{\ss}lin, Lieu, \& Biermann (l999) have explored processes that might
produce a heretofore undetected population of lower energy cosmic rays which
could produce this flux. They demonstrate an evolutionary scenario in which
relativistic electrons produced in the last merger event in Coma two Gyrs ago
could produce these electrons.  However, this model cannot produce the spatial
profile of the EUV emission obtained by Bowyer \& Bergh\"ofer. They also
consider IC scattering of starlight photons and show that under some scenarios
this could account for the EUV flux and the required spatial distribution.

Sarazin \& Lieu (1998) have suggested that all clusters of galaxies may
contain a relic population of cosmic ray electrons that are unobservable in
the radio and these will produce excess EUV flux by inverse Compton scattering
against the 3$^{\circ}$\,K cosmic background. Their proposal is based upon,
and explains, details of the EUV data obtained on Abell 1795 (Mittaz, Lieu \&
Lockman, 1998).

We have obtained new data on some of these clusters and have analyzed archival
data on others. We find that the results obtained are crucially dependent upon 
the characterization of the DS telescope, and upon details of the estimation of
the EUV emission from the X-ray plasma. The results we obtain are quite
different from those obtained in previous work.

\section{Data and Data Analysis}  

All data were processed using procedures of the IRAF EUV package provided by
the Center for EUV Astrophysics (CEA, Berkeley) which were especially designed
for the analysis of EUVE data. As part of this process, we excluded detector
events with pulse heights far from Gaussian peak of the photon pulse-height
spectrum. Low energy events due to spurious detector noise($\approx$15\% of
the total), and high energy counts due to cosmic rays and charged particles
($\approx$25\% of the total), were screened out. A detailed description of
different background contributions to the DS data can be found in Bergh\"ofer
et al. (1998). We point out that the location of the Gaussian peak in the
pulse-height spectrum is not constant for all EUVE DS observations since the
gain of the DS detector was changed periodically in the course of the mission.
Consequently, pulse height limits were chosen individually for each DS
observation.The resulting filtered event lists were corrected for electronic
deadtime and telemetry throughput effects.

The background of the DS telescope consists of a uniform detector background,
$B_{\rm int}$, and a component that may vary over the field because of a
variety of effects including vignetting, variations in the thickness of the
filter covering the detector face, variations in the quantum efficiency over
the face of the detector, and other causes. Hereafter we call this second
component the vignetted background, $B_{\rm vig}$. To investigate the
possibility of a field variation effect, we chose four 20,000 s observations
of blank sky with low and similar backgrounds that were obtained in a search
for EUV emission from nearby pulsars (Korpela \& Bowyer, 1998). We added
90,000 s of data from a blank field at R.\,A.$_{2000} = 3^h31^m39^s$ ,
Dec.$_{2000} = +18^{\circ}28\arcmin33\arcsec$ obtained from the EUV
archives. {\em We emphasize that the images were added in detector
coordinates, NOT sky map coordinates.}
We processed the data as described above. We established that once
proper pulse height selection of the detector events had been made, the
detector backgrounds were all spatially identical.

\begin{figure}
 \resizebox{\hsize}{!}{\includegraphics{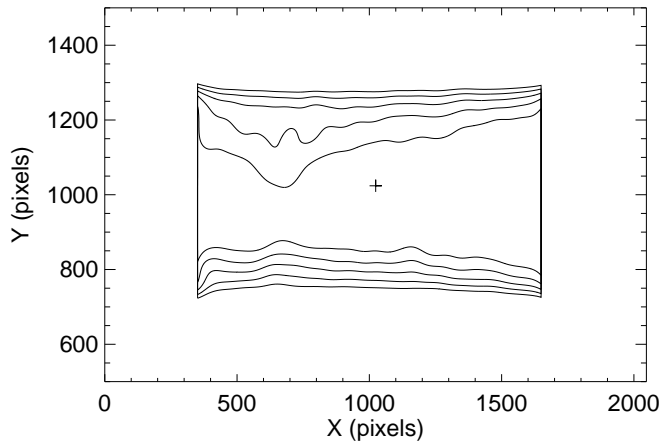}}
\caption[]{\label{sensi} A contour plot of counts obtained in long duration DS
  exposures showing the sensitivity variation of the DS Telescope over the
  field of view. We have cut the regions at the detector ends where the
  detector distortion becomes severe. The field displayed is approximately
  1.75 degrees x 0.73 degrees.}
\end{figure}

A contour plot of normalized count rates in these exposures convolved with a
32 pixel wide Gaussian is shown in Figure\ \ref{sensi}. The contours represent 
a 10\% change in the measured count rates.  \footnote{Investigators interested
in using this observationally derived sensitivity plot may access Fits Files at
  ``http://sag-www.ssl.berkeley.edu/$\sim$korpela/euve\_eff'' .}  It is
informative to compare this observationally derived result with the
theoretically derived product provided in Sirk et al. (1997), which has been
used in previous work on EUV emission in clusters and is essentially flat.

\begin{figure}
 \resizebox{\hsize}{!}{\includegraphics{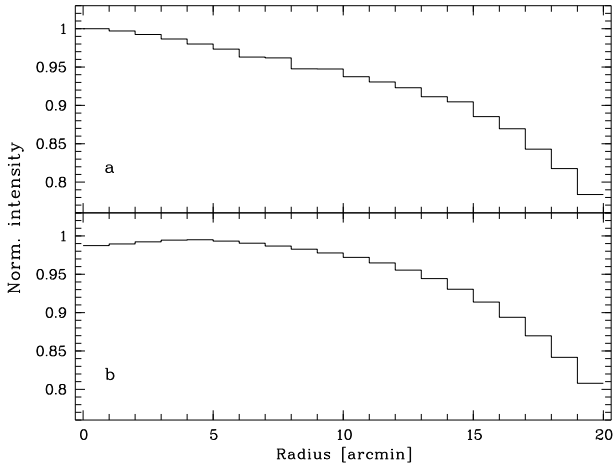}}
\caption[]{\label{bprofile} a: The radial intensity profile of the EUV 
emission from the blank field shown in Fig.\ \ref{sensi} with the assumption 
that any underlying background is flat.  Fig.\ \ref{bprofile}a shows the 
profile 15\arcmin\ to the left of the boresight and Fig.\ \ref{bprofile}b
shows the profile 2\arcmin\ to the right of the boresight. Both profiles show
(false) diffuse emission; each is different because of differing telescope
sensitivity variations at these locations.}
\end{figure}

As a demonstration of the effect of this variation of telescope sensitivity
in regards to studies of diffuse emission, we have derived azimuthally 
averaged radial intensity profiles of the EUV emission of the blank field
shown in Figure\ \ref{sensi} under the assumption that any background present 
was flat, following the procedures of Lieu et al. (1996) and Mittaz et al. 
(1998).  One of these profiles was centered 15\arcmin\ to the left of the 
boresight, and one was centered 2\arcmin\ to the right of the boresight, 
following observation strategies often employed with EUVE.  These results are 
shown as Figures\ \ref{bprofile}a and \ref{bprofile}b.  These
profiles clearly show (false) extended emission centered at these locations.
A still different radial profile would be obtained at different locations on
the detector and still different profiles would be obtained if data from any
two locations were added.

In order to account for this telescope sensitivity variation in searching for
true diffuse emission in clusters of galaxies, one must carry out detailed,
though straightforward, processes.

All observations contain both $B_{\rm int}$ and $B_{\rm vig}$. Because the
ratio of these two backgrounds can vary, one must correct for this effect
when scaling previously measured backgrounds to the backgrounds of the new
observations. The background subtracted image is:

\begin{equation}
I_{\rm net} = I_{\rm on} - B_{\rm int} - f B_{\rm vig} 
\end{equation}

where $I_{\rm on}$ is the on-source image.  The term $B_{\rm int}$ is derived
from measurements of the background in obscured regions of the detector
covering about 3.5 \% of the detector area.  The term $B_{\rm vig}$
represents the vignetted background.  The factor $f$ is used to fit the 
vignetted background levels in the blank field with those of the on-source 
image. This factor is derived by fitting the observed photonic background with
that of the blank field images in a region far from the source.
Because of the long duration of the background exposures, the
statistical errors in $f$ are less than 1\%.  When comparing on-source and
background in small detector regions, our errors are dominated by the count
statistics of the region, rather than errors in the background fitting.

We have examined new data on Abell 1795 with the archival data on this
cluster and find that the raw data from the new observations at $R >
2\arcmin$ (which excludes the effects of a bright transient source in the new 
data set) are identical within the counting errors, confirming the validity of
the original data set used by Mittaz et al. (1998). Because the two data sets
are identical and the more recent set is contaminated with a point source, we 
have used the archival data on Abell 1795 for our subsequent analysis.

\begin{figure}
 \resizebox{\hsize}{!}{\includegraphics{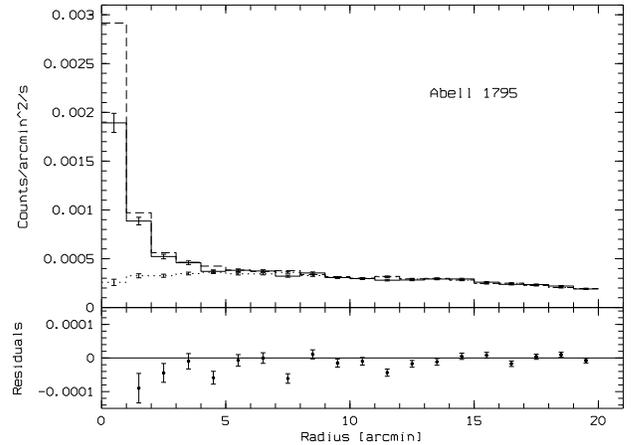}}
\caption[]{\label{rad1795} The azimuthally averaged radial intensity profile of
  the EUV flux in Abell 1795 is shown as a solid line. The vignetted
  background from long observations of blank fields is shown as a dotted line.
  There is no excess EUV emission beyond 4\arcmin.}
\end{figure}

We derived the azimuthally averaged radial intensity profile of the EUV
emission of the cluster as a function of projected radius from the central
core assuming spherical symmetry. The results are shown in Figure\
\ref{rad1795}. Our vignetted background, fitted at $R > 15\arcmin$, is shown 
as a
dotted line. It is visually apparent that there is no excess EUV emission at 
radii larger than 4\arcmin. It is also clear that an improperly chosen 
background chosen at $R > 15\arcmin$ would result in apparent emission at 
smaller radii simply because of the effects of the vignetted background in the
DS telescope.

We next determined the expected intensity and distribution of the EUV emission
expected from the X-ray emitting plasma. We used the X-ray radial emission
profile provided by Briel \& Henry (1996). This profile was derived from ROSAT
PSPC observations of the cluster in the energy band between 0.5--2.4 keV. At
larger radii $(R > 4 \arcmin)$ this profile is well fit by a King profile
(1972) with $\beta = 0.93$ and describes the large scale cluster X-ray
emission with a temperature of 6.7 keV. The ROSAT observations also show a
central excess emission within $R < 4\arcmin$.  Briel \& Henry (1996) obtained
a temperature of 2.9 keV for this excess. We derived conversion factors for
counts in the 0.5--2.4 keV band of the ROSAT PSPC to EUVE DS counts using these
plasma temperatures. Our derivation employed the MEKAL plasma emission code 
with abundances of 0.3 solar. For a temperature of 6.7 keV we obtained a 
conversion factor of 126; the value for 2.9 keV was 110. We found that varying
the temperature by $\pm$1 keV or using different abundances only affect these
conversion factors by a few percent and thus changes of this nature would not
significantly alter our results. We found that a deprojection of the emission 
components which takes into account the emission measures and sizes of the
different components leads to the same result.

The correction for the intervening absorption of the ISM in our Galaxy will
have a substantial impact on our results. Many workers simply apply the cross
sections of Morrison \& McCammon (1983) or Baluci\'nska-Church \& McCammon
(1992) for this correction, but there are several problems with this approach.
The \HeI absorption coefficient in
this work is incorrect  (Arabadjis \& Bregman 1999). 
In addition, the ionization state of the ISM will substantially affect the
result.  The ISM absorption at EUV energies is primarily due to \HI, \HeI,
and \HeII; the metals in the list of Baluci\'nska-Church \& McCammon (1992) 
provide
less than 30\% of the absorption at wavelengths greater than 50 \angstrom, and
less than 10\% at wavelengths greater than 100 \angstrom, and none of the 
Galactic ISM is in the form of \HeIII (see discussion below).  
Hence in general terms the absorbing material and related factors are given by:
\begin{equation}
N({\rm H(tot)}) = N(\HI) + N(\HII)
\end{equation}
\begin{equation}
N(\HeI) = 
    \frac{1}{10}\left[ N({\rm H(tot)}) \right] \left( 1 - X(\HeII) \right)
\end{equation}
\begin{equation}
N(\HeII) = 
    \frac{1}{10}\left[ N({\rm H(tot)}) \right] \left( X(\HeII) \right) \\
\end{equation}

We have calculated the Galactic ISM absorption using these columns with \HI
cross sections of Rumph \etal (1994), \HeI
cross sections from Yan \etal (1998), and \HeII cross sections from Rumph \etal
We used these values with an improved estimate of the Galactic
neutral hydrogen column density in the direction of Abell 1795 of N(H{\sc I})
= $0.95 \times 10^{20}$cm$^{-2}$ (J. Lockman, private communication) .
We assume the total helium is 10\% of the total hydrogen column.
A direct measurement of the \HII column can be obtained, in principle, from
measurements of the \Halpha flux in this direction (Reynolds \etal, 1998).
Unfortunately, only an upper limit to this flux of $1 \times 10^{19} 
{\rm cm}^{-2}$ is currently available (Haffner, private communication).
A reasonable estimate for the \HII column, based on all the available data, is 
that it is close to this upper limit (Reynolds, private communication).
Consequently we have used this value for the \HII column.  The
amount of \HeII in this direction can be obtained from Fig.\ 1 of Bowyer
\etal (1996).  
For A1795, $N({\rm H(tot)}) = 1.1 N(\HI)$,
$N(\HeI) = 0.1 [1.1 N(\HI)](1-0.02)=0.108 N(\HI)$, 
and $N(\HeII) = 0.1 (1.1 N(\HI)] \times 0.02= 2.2\times 10^{-3} N(\HI)$.
The absorption corrected results are shown in Figure\ \ref{rad1795} as a dashed
line. The observed EUV emission is {\it less} than that produced by the X-ray 
plasma. This appears to be unphysical but is simply understood as discussed 
below.

After taking into account the vignetting and the EUV emission from the X-ray
plasma, we see no excess EUV emission in this cluster.

\begin{figure}
 \resizebox{\hsize}{!}{\includegraphics{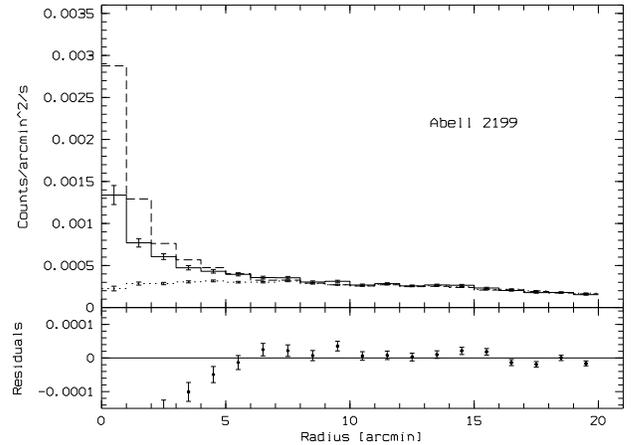}}
\caption[]{\label{rad2199} The azimuthally averaged radial intensity profile of
 the EUV emission in Abell 2199 is shown as a solid line. The dotted line is 
 the vignetted background. There is no EUV emission beyond 8\arcmin.}
\end{figure}

We also examined archival data on Abell 2199 to ascertain whether a vignetted
background could have produced an artificial extended diffuse EUV halo in this
cluster. In Figure\ \ref{rad2199} we show the radial profile of the raw EUV 
data and the vignetted background fitted at $R > 15 \arcmin$. 
It is apparent that there is no excess EUV emission
beyond 8\arcmin. We use the results of Siddiqui, Stewart, and Johnstone,
(1998) to model the EUV emission from the X-ray gas in the cluster. They found
T(core) = 2.9 keV and T(outer) = 4.08 keV. The conversion of the ROSAT
X-ray profile into EUVE DS count rates has been done as described for Abell 
1795. For Abell 2199 we found DS to PSPC hard band count rate ratios of 
83.2 for T = 2.9 keV and 89.4 for T = 4.08 keV. Absorption by the Galactic ISM 
was accounted for using $N(HI)= 8.3 \times 10^{19}$ (Lieu et al. 1999) with 
ionization fractions and cross sections as described previously. The results 
are shown in Figure\ \ref{rad2199} as a dashed line. Again, the expected EUV 
emission from the X-ray gas is larger than the observed flux, and there is no
excess EUV emission in this cluster.

\begin{figure}
 \resizebox{\hsize}{!}{\includegraphics{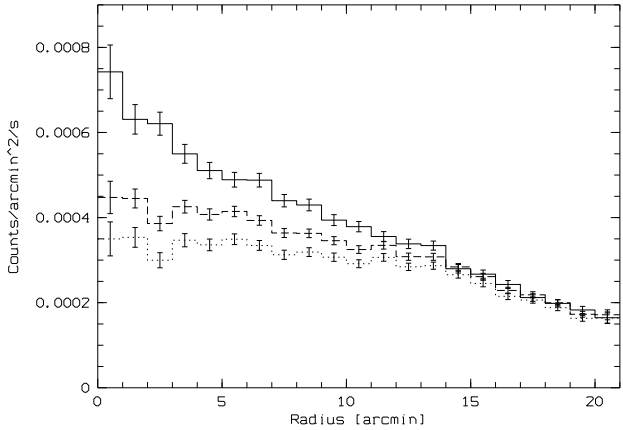}}
\caption[]{\label{radcoma} The azimuthally averaged radial intensity profile of
  the EUV flux in the Coma cluster is shown as a solid line. The expected EUV
  emission from the X-ray plasma is shown as a dashed line. The vignetted
  background is shown as a dotted line.}
\end{figure}

Finally, we re-examined the previously reported
EUV excess in the Coma Cluster.  We carried out our analysis using both of the
existing DS images of this cluster.  Because of the different roll orientation
and pointing position in these images, it was necessary to carry out our
analysis on each image individually.  The results were then summed and the EUV
emission and vignetted background are shown in Figure\ \ref{radcoma} as a solid
and dotted
line respectively. In this figure, we have fit the vignetted background to the
Coma observations beyond 17\arcmin; however, because faint emission due to the
cluster likely extends past this point, especially in the direction of the NGC
4874 subcluster, this is likely to be a slight overestimate of the background
and hence the excess EUV emission we derive may be a slight underestimate. If
the X-ray profile of the Coma Cluster is used as a guide, we expect this
effect to be small compared to the statistical errors in each radial bin.

The X-ray profile has been constructed using ROSAT PSPC archival data of Coma.
We verified that our PSPC hard band cluster profile is consistent with the
profile provided by Briel, Henry \& B\"ohringer (1992) but includes the
central excess associated with the galaxy group around NGC 4874. We assumed
that this X-ray emission is due to a plasma at T = 9 keV (Donnelly et al.,
1999) absorbed by a hydrogen column of $8.7 \times 10^{19} {\rm cm}^{-2}$
(Lieu et al., 1996) with ionization fractions and cross sections for 
Galactic absorption as described above. Here we obtained a DS to PSPC hard 
band conversion factor of 112.

\begin{figure}
 \resizebox{\hsize}{!}{\includegraphics{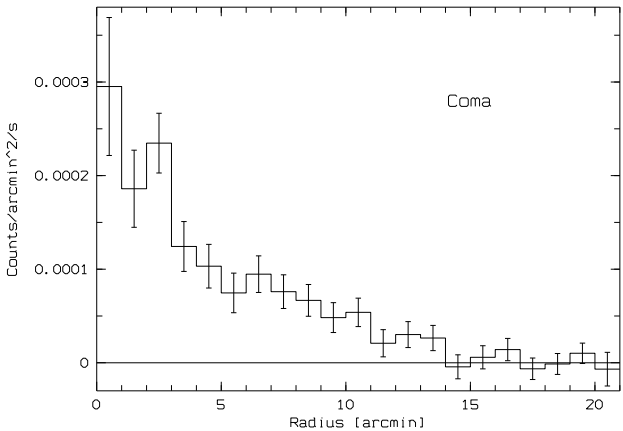}}
\caption[]{\label{excoma} The excess EUV emission in the Coma cluster.}
\end{figure}

The residual EUV emission in excess of the expected contribution of the X-ray
gas shown in Figure\ \ref{excoma} demonstrates that there is, indeed, excess 
EUV emission in the Coma cluster. 

\begin{figure}
 \resizebox{\hsize}{!}{\includegraphics{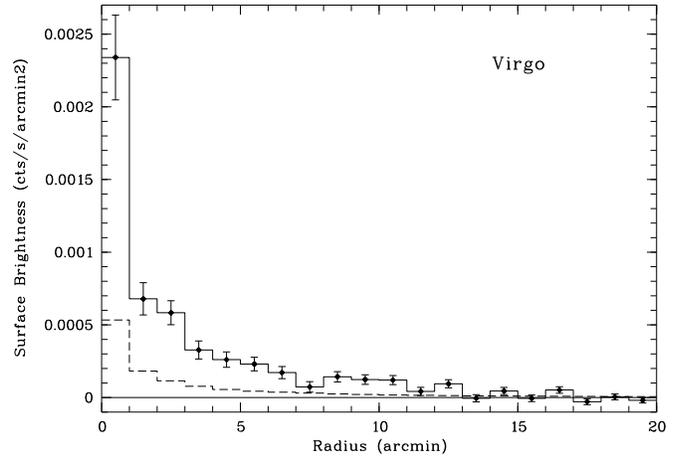}}
\caption[]{\label{exvirgo} The excess EUV emission in the center of the Virgo 
cluster (solid line) in comparison to the expected contribution of the X-ray 
emitting cluster gas (dashed line).}
\end{figure}

Based on our new analysis technique we also confirm excess EUV emission in the
Virgo cluster of galaxies. Details are provided in Bergh\"ofer \& Bowyer 
(this workshop). In Figure\ \ref{exvirgo} we provide a plot of the excess EUV 
emission in the central part of the Virgo cluster. Clearly, there is EUV 
emission in excess of the expected low energy tail of the X-ray emitting gas 
in Virgo.

\section{Discussion}

The results of our new analysis show no excess EUV emission at radii larger
than 4\arcmin\ for Abell 1795 and 8\arcmin\ for Abell 2199
contrary to previous work on these clusters. When we consider the
inner regions for these clusters, we find the results expected are dependent
upon a proper evaluation of the EUV emission from the X-ray plasma.  When this
emission is properly accounted for, the expected EUV emission from the
X-ray plasma is {\it less} than is actually produced.
This can be understood in terms of excess absorption within the cluster core.
This effect has been noted in studies of X-ray cluster emission in cooling
flows, where it is often reported that the hydrogen column density is larger
in the core of the cluster. There is no observational evidence for more
hydrogen in these regions, and neutral hydrogen is not expected in this
environment. A reasonable explanation for this effect is that the X-ray
reduction codes employed in these analyses require more absorption for a
reasonable fit, and this is achieved blindly by adding more hydrogen with a
standard admixture of non-ionized metals. It is more likely that in the cooler
regions of the cooling flow, some metals are not completely ionized and these
ions produce the extra absorption of the X-ray flux (Allen et al. 1996).
This absorption would be even more substantial for the EUV flux, and would
produce the effects seen. We point out that a study of the
differing amount of absorption in the EUV and X-ray bands may provide
sufficient information to identify the primary absorbing species.

When we employ our new analysis techniques with the data on the Coma cluster
we find there {\it is} excess EUV emission confirming the results of previous
studies. However, the distribution of this flux differs in detail from that
previously reported. The distribution of this radiation is shown in Figure\ 
\ref{excoma},
along with the count rate intensity. The intensity in physical units is 
(slightly) dependent upon the assumed spectral distribution of the flux. A 
source with a photon spectral index of 1.6 results in an EUV source luminosity
of $1.5 \pm 0.5 \times 10^{42}$erg\,s$^{-1}$.

It is useful to consider why our results are different from those of Mittaz et
al. (1998), and Lieu, Bonamente \& Mittaz (1999).  While it is difficult to 
evaluate the details of another researchers' analysis, it is clear that a key 
difference is our use of an observationally derived vignetted background. 
Mittaz et al. and Lieu et al. used the theoretical background function 
(Richard Lieu, private communication) which is essentially flat 
In addition, these authors also carried out their analysis of the EUV flux 
without first removing the non-photonic background from their data.  
The extent to which this affects the results is not substantial, as we obtain 
the same general picture with data that has not been processed in this manner.
Their approach to estimating the EUV emission produced by the
X-ray plasma is also different than ours. 

It is interesting to ask why both
the analyses of Lieu et al and that presented here do show excess diffuse EUV 
emission in the Coma cluster and the Virgo cluster. The primary
explanation is that both of these clusters do, in fact, have excess EUV 
emission. This emission is sufficiently extended that the effects of the 
vignetted background, though changing the details of the results, do not
dominate as they do in Abell 1795 and Abell 2199.

\begin{figure}
 \resizebox{\hsize}{!}{\includegraphics{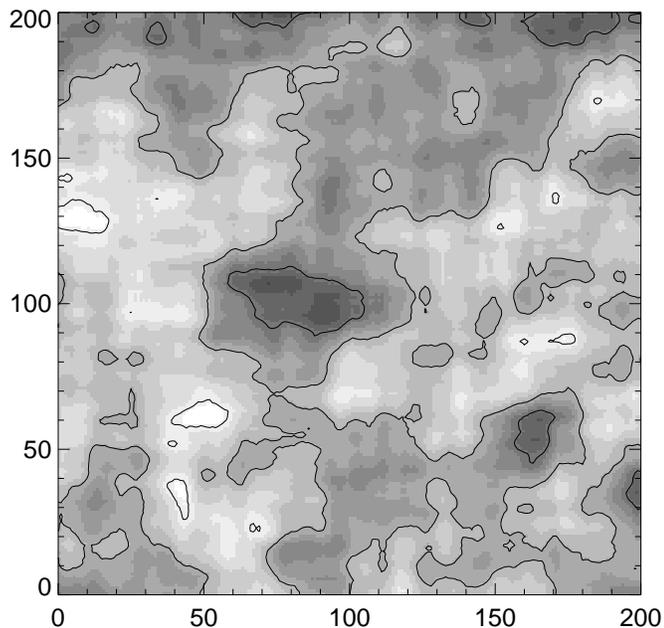}}
\caption[]{\label{sensi2} Variations in the blank field 
image for a 15\arcmin\ field of view centered upon the detector coordinates
where the EUVE observation of A 2199 was made.  The background was smoothed
to 2\arcmin.  Gaussian count statistics give expected uncertainties
of  2.6\%.  The contours are 90\%, 100\%, and 110\% of the mean and are each 
separated by 3.8$\sigma$.}
\end{figure}
  
Finally, we have examined the effects of fine grained detector variations on
the data obtained from any point in the sky.  As one example in Fig.\ \ref{sensi2}
we show variations in the blank field image for a 15\arcmin\ field of view
centered upon the detector coordinates where the EUVE observation of A 2199
was made.  The image has been smoothed to 2\arcmin.  Count statistics give
expected uncertainties of 2.6\%.  The contours are 90\%, 100\% and 110\% of the
mean and are separated by 3.8 $\sigma$.
A wavelet analysis of this data as presented by Richard Lieu (this conference)
would obviously show substantial effects, which would be entirely due to
this fine scale structure in the detector.  This structure varies over the
face of the detector.  Consequently the image taken at different locations
will be different, which will tend to confuse uncareful workers.

\section{Conclusions}
We investigated the effects of the telescope sensitivity variation over the
field of view and found this was a key factor in investigating extended EUV 
emission in clusters of galaxies. 
Our study shows why excess EUV emission has been found in
every cluster examined to date with EUVE. {\it Any} point in the sky will show
extended EUV emission using the analysis employed in previous studies of
clusters of galaxies.  We also used a detailed approach to the 
evaluation of the EUV flux produced by the X-ray gas in the core regions of 
these clusters. 

We find no evidence for excess EUV emission in Abell 1795 or Abell
2199. We do, however, confirm extended EUV emission in the Coma cluster and
Virgo cluster although the distributions of these fluxes are different in 
detail from that previously reported. The fact that we do find extended EUV 
emission in the Coma cluster and Virgo cluster using our new analysis 
procedures confirms that an unidentified processes is operative in this 
cluster.

\begin{acknowledgements}
We acknowledge useful discussions with Michael Lampton, Pat Henry, John
Vallerga, Carl Heiles, Richard Lieu, John Mittaz and Jean Dupuis. This work
was supported in part by NASA contract NAS 5-30180. TWB was supported in part
by a Feodor-Lynen Fellowship of the Alexander-von-Humboldt-Stiftung.
\end{acknowledgements}

\end{document}